\documentclass[preprint]{elsarticle}
\usepackage{color}
\usepackage{graphicx}
\usepackage{url}
\begin{document}

\title{cuInspiral: prototype gravitational waves detection pipeline fully coded on GPU using CUDA}           
\author[Bosi]{Leone B. Bosi\corref{cor1}\fnref{fn11}}        
\ead{leone.bosi@pg.infn.it}
\cortext[cor1]{Corresponding author}
\address[Bosi]{INFN - Via Pascoli, 1 Perugia Italy}
\begin{abstract}
In this paper we report the prototype of the first coalescing binary detection pipeline fully implemented on NVIDIA GPU hardware accelerators. The code has been embedded in a GPU library, called cuInspiral and has been developed under CUDA framework. The library contains for example a PN gravitational wave signal generator, matched filtering/FFT and detection algorithms that have been profiled and compared with the corresponding CPU code with dedicated benchmark in order to provide gain factor respect to the standard CPU implementation. In the paper we present performances and accuracy results about some of the main important elements of the pipeline, demonstrating the feasibility and the chance of obtain an impressive computing gain from these new many-core architectures in the perspective of the second and third generations of gravitational wave detectors.    
\end{abstract}     

\maketitle

\include{introduction}
Recently, innovative solutions have been implemented by CPU producers, introducing highly parallel design in processors architecture. The productive process is evolving  from the "multi-core" era toward the "many-core" era, where hundreds of thousands of processing core are embedded on the same processor. Such kind of changes on architectural paradigm require an identical change on the programming paradigm. Modern, GPUs(Graphics Processing Unit) fall in the manycore definition, being characterized by hundreds of processing cores. Moreover due to the inherently parallel nature of 2D and 3D computer graphics problems, GPUs are well suited for more general problems, performing complex computation in very general fields. Since an impressive develop is taking place in the GPU technology, it makes sense to explore GPU performances not only for graphics, but also for other applications. The achievable speedup on GPU is strongly correlated to the algorithms implementation, which  can be more than  2 orders of magnitude under optimal conditions. Therefore we can claim that a so big improvements in performances has an effect also on how science is made, giving access to new solutions that were unaccessible or very difficult to achieve before.

There are some important news also under the GPU devices programming simplification and unification. On that we have to remark  CUDA architecture\cite{CUDA} developed by NVIDIA and OpenCL\cite{OpenCL}, developed inside the Khronos Consortium. This last project defines a common language for parallel computing devices and it is supported by the major semiconductor and hardware 
vendors. The work shown in this paper has been developed on CUDA.
\\
\\
In this paper we show the implementation of these technologies in the  context of the gravitational wave astronomy.  In particular we present the first prototype of a detection pipeline for coalescing binaries (CB) gravitational wave signal completely GPU code based.

Actually the two most important  ground-based gravitational-wave (GW) detectors had completed long science data taking, with a very impressive sensitivity close to their design performance\cite{VIRGOLIGOrun1,VIRGOLIGOrun2}. These are Virgo\cite{VIRGOref1,VIRGOref2} in Italy and LIGO\cite{LIGOref} in United States. During the next years this first generation of detectors will undergo to several upgrades, moving toward enhanced and advanced versions. 
The direct detection of the first GW signal is expected to be made by Advanced versions, where the CB events rate is enough to have an high degree of confidence. About third generation, in 2009,  the Einstein Telescope\cite{ETref} project stared. This is a design study about detectors of third generation and is supported by the European Commission under the Framework Program 7 (FP7, Grant Agreement 211743). It concerns the study and the conceptual design for a new research infrastructure that will bring Europe to the forefront of the most promising new development in our quest to understand the history and future of the Universe, the emergence of the field of Gravitational Wave Astronomy\cite{ETfuture,ETfuture1}. The gain in terms of sensibility between first ground based gw detector and ET like detectors is expected to be 2 order of magnitude more, increasing exponentially the demanded data analysis computing power. 
\\
\\
One of the most promising sources of gravitational wave signal for detection purpose are the coalescing  binaries systems, which consist on  two fast rotating compact objects, such as neutron stars and black holes, that loses orbital energy under gravitational wave emission\cite{HulseTaylor}.

The algorithm used for the optimal detection procedure is based on matched filtering (MF) technique\cite{MF1,MF2}, where the detector signal is "compared" via correlation with a set of expected theoretical signals, called templates.  This technique is very computing demanding, because for a complete analysis an high number of templates(the MF filters) have to be considered  and moreover because the core algorithm is the Fast Fourier Transform(FFT)\cite{FFT1}. 

The templates number is a critical issue, depending strongly on the detector sensitivity and on the analysis accuracy. Actually we are facing the detection procedure using roughly some thousands of templates. This amount is enough for the first generation detectors, but if we consider ET generation detectors, we have estimated a number of 1-2 million of templates to perform the same detection strategy, increasing the complexity of some orders of magnitude. \footnote{At such scale the brute force is not the best solution, and the procedure can be made more skilled, permitting to speedup the analysis.} 

In this paper we report the first coalescing binary detection pipeline, composed by template bank generation, matched filtering procedure and maxim estimation 
 based completely on GPU. In the next section we describe this items with more details. 

With this work we demonstrate the feasibility to use these architectures for gravitational wave physics purpose, obtaining a factor 50-100 on the actual performances and the applicability to cover computational problems related to the detectors of third generation as ET with the perspective of the manycore computing era.

\section{Graphics Processing Processor(GPU) and CUDA}

The prototype pipeline presented in this work has been developed using CUDA, the computing engine of the NVIDIA Processors. The code has been written programming with "C for CUDA", the specific environment to compile code for NVIDIA GPU. In this framework the programmer can define special C functions, called kernels, 
that, when called, are executed N times in parallel by N different CUDA threads into GPU core processors. In some sense this architecture is an extension of SIMD model. In detail, when a kernel is executed, N copies of it are generated and executed in parallel inside GPU device; moreover at each thread a unique id is assigned, starting from $0$ to $N-1$, which can be used as unique identifier inside the kernel code for several purpose, such as accessing memory. 

In the CUDA programming model, CUDA threads execute on a physically separate device, operating in some sense  as a co\-processor, respect to host computer
that runs the C program. Sometime this programming model is reported as SIMT(Single Data Multiple Threads) as likeliness to SIMD, where instead of single data multiple data, a single kernel execution produces a multiple pool of threads running.

CUDA has some advantages respect to other general purpose computation on GPUs (GPGPU)\cite{GPGPU}:
\begin{itemize}
\item Scattered reads – code can read from arbitrary addresses in memory.
\item Shared memory - this can be shared among threads and can be used as a user-managed cache, and permitting higher bandwidth.  
\item L1 and L2 cache on Fermi architecture
\item Faster downloads and read backs to and from the GPU 
\item Full support for integer and bitwise operations, including integer texture lookups. 
\item Very good support and SDK
\end{itemize}

The code presented in this paper has been tested on NVIDIA GTX275 board (GT200 series), characterized by 240 computing core, GDDR3 160 GB/s Memory, 1TFlops peak performance and a consumption of 200W. 

\section{Coalescing binary signal detection problem}

In this section a brief introduction to the gravitational wave signal characteristics and detection procedures are reported.

Gravitational waves have been predicted by A. Einstein in 1916 as general relativity theory implication\cite{GWEinstein}. Under particular circumstances, accelerated masses could produce a perturbation in the space-time that propagates as a wave. Gravitational radiation has not yet been directly detected, but there are many indirect proofs of its existence. it was remarkable the Nobel Prize in Physics, awarded for measurements of the Hulse-Taylor binary system in 1993\cite{HulseTaylor}. 

Sources of gravitational waves include binary star systems composed by compact objects like Neutron Stars or Black Holes. In these systems orbital energy is carried out via gravitational wave, producing a progressive reduction of the relative distance between the two stars. This process continues since the coalescence event, where the two starts stop to live as distinct objects, beginning to merge. Coalescing compact binaries with neutron star or black hole components provide the most promising sources of gravitational radiation for detection by the interferometric detectors experiments.

Gravitational waves (GW) are transverse waves, which effect on metric distortion  perpendicularly to the propagation direction. A GW can be expressed by \textbf{two} polarizations:
$\mathbf{h(t)}=\alpha \mathbf{h_{+}(t)}+\beta \mathbf{h_{\times}(t)}$. The prediction about the signal shape can be made theoretically, but, due to the complexity of the general relativity equations,  it can be made only with approximation methods. The expression describing the signal is function of a set of parameters, which depend on source physical characteristics\cite{GWapprox2,GWapprox3}. A typical set of parameters is $(m_1,m_2, R, i, \phi)$, where:
\begin{itemize}
\item m1;m2 - star masses
\item R - relative distance between the binary system and the observer
\item i - inclination angle
\item $\phi$ - initial phase
\item $t_0$ - arrival time
\end{itemize}
\subsection{Matched filtering}
On defining the detection strategy a big lack is about signal knowledge, because we  don't know which and when the binary system will emit a GW.   

Under the conditions of two polarization signal with a set of unknown parameters, the detection theory of signal provides a procedure based on  Neyman-Pearson method, permitting the definition of the optimal detector. In this case the "detection rule" is based on matched filtering technique\cite{MF2}, that is the optimal linear filter that maximizing the \textit{signal-to-noise-ratio} (SNR) in presence of additive stochastic noise. 

A matched filtering between two signals is obtained correlating a \textit{template} of the signal that we are looking for, used as reference signal, with the stream of data where is assumed to be present the signal itself. This operation is equivalent to convolve the input data stream with a conjugated time-reversed version of the template (cross-correlation). 

In term of computational complexity, we use to apply matched filtering  in frequency domain, where it is possible to benefit of the FFT algorithm\cite{FFT1}, having  a complexity of $O(N\log(N))$, where $N$ is the length of the two vectors of data and transforming the time correlation in a complex vector product. 

Technically, given the input stream of data $x(t)$,  the unknown signal $\theta(t)$ that we are looking for and the power spectrum of the noise $S(f)$, we can define the matched filter in presence of colored noise as:
\[\label{eq:matchedfilter}c(t)=\int \frac{X(f)\Theta^{*}(f)}{S_n(f)}e^{ 2\pi ift}df \equiv<x(t),\theta(t)>\] .This formula can be use to define the matched filter as an inner product.

In case of Gaussian-white noise, the previous expression is reduced to the standard correlation function \cite{MF2}. For that the \ref{eq:matchedfilter} is sometime referred as correlator.

Due to what has been previously reported, the set of parameters  $(m1,m2,R,i,\phi)$ describing the system is completely unknown. Under these hypothesis the detection procedure is completely blind and require to use a set of templates, which are built a priori. Each template is associated with a point in the masses parameters space $(m1,m2)_i$, describing the signal sources, and the parameter space is defined in order to cover the region where signal is expected to be. The template placement is sampled in such a way to have a tolerable mismatch error between close templates\footnote{Usually a 2-5\% of mismatch error is used}. The number and density of points, composing the grid, has to be enough and correctly distributed in order to reduce the possibility of a signal detection miss, due to the mismatch between the real signal and the templates bank filters. 

The procedure\cite{CBDET1,CBDET2,CBDET3} needs to compare each grid point, corresponding to a specific template $i$/$(m1,m2)_i$, with the data stream using the matched filtering and than to look for the values above a certain predefined threshold, which depends on detection probability and the a priori false alarm decision\cite{MF2}.
\\
\\
We can synthesize the pipeline steps for each input data chunk, as follow:\\
- Loop over template bank, where for each parameters couple $(m1_i,m2_i)$:
\begin{enumerate}
	\item generate and normalize the template $\theta(t,m1_i,m2_i)$ in order to be $<\theta(t),\theta(t)>=1$
	\item calculate the two correlators with matched filter for the two polarizations: $c_+^i(t)=<x(t),\theta_+(t)>$,$c_\times(t)=<x(t),\theta_\times(t)>$
	\item add two polarization in quadrature, $c^i(t)=\sqrt{c_+^i(t)^2+c_\times^i(t)^2}$ 
	\item look for $\{t_0^i\} =max(c^i(t)>\lambda)$, where $\lambda$ is a predefined threshold. The set ${t_0^i}$ is the ensemble of the candidate arrival time. The amplitudes $c(t_0^i)$ give the signal-to-noise ratio, proportional to the detection probability.
\end{enumerate}

- Reduce the [$c^i(t_0^i)$] set, in order to detect local maximum

\section{Pipeline Implementation}
The prototype pipeline for coalescing binary signal, reported in this paper, is based on the procedure described in the previous section. In detail the code loads the template bank data from an ASCII file and than it follows the detection procedure as reported  in figure \ref{fig:pipeline}.
In order to obtain the maximum benefit on using GPU, it fundamental to 
make all possible computation on GPU board, reducing or hiding\footnote{This means to make, if it is possible, data IO during GPU computation. This permits to hide data transfer weight to the global computation time.} as much as possible  data exchange between GPU memory and PC main memory and  increasing as much as possible the GPU computational density. 
By these consideration we can conclude that the approach where the GPU is used just as a simple co\-processor it is not the best one (for example wrapping just some kernel\/functions, such as FFT\cite{GPUCBwrong}), because the host/device IO can dramatically  reduce performance  respect to what these new technologies can express. For that, with the purpose of exploring such new multicore architecture, we have developed \textsl{cuInspiral}, a prototype library, containing functions for the CB detection pipeline fully working on GPU device.

In our work the pipeline has been developed defining the GPU library with a set of CUDA kernels and host functions which execute the CB pipeline algorithms all inside GPU device, except for some "reduction" algorithms that need  CPU contribute. In this section we report about this CB signal detection code based fully on GPU, showing the achieved speed up respect to similar CPU implementation. 

\subsection{Pipeline main details}
\begin{figure}[t]
\centering
\includegraphics[width=4in]{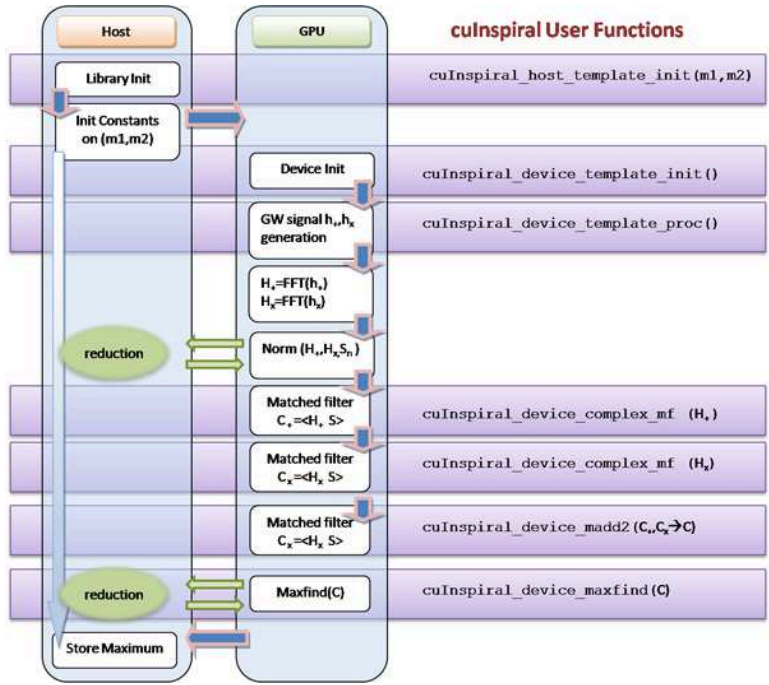}
\caption{In this plto we report the rough design of cuInspiral pipeline and the user level functions involved in the computation On the left side is represented  the host computing side and on the right the GPU computing side. The pipeline in mainly characterized by template generation, normalization, FFT, matched filtering product in frequency domain, iFFT, events detection and reduction}
\label{fig:pipeline}
\end{figure}

The core of the analysis is made in frequency domain, in order to get vantage from the DFT algorithm. In detail we use  CUFFT library, the FFT implementation for nvidia GPU\cite{CUDAsdk}. In figure \ref{fig:pipeline} we report a schema of the CB pipeline, dividing the data/work flow between  host part (left side) and GPU part (right side). In the same picture we report also the name of the library functions involved at each step.
In order to estimate GPU/CPU speed\-up, we have created a CPU counterpart code for some of the GPU routines. This choice permits us to perform  performances and accuracy analysis, using the CPU results as reference.

Here we report a brief description about some of the pipeline functions, as shown in  \ref{fig:pipeline}:

\begin{description}
    \item[host\_template\_init] this function acts on host side and it initializes the generator constants, starting from  star masses $(m1,m2)$ of the template bank, passed as arguments.
    \item[device\_template\_init] initialized constants are passed to the GPU memory.
    \item[device\_template\_proc] this command calls the generator kernel, filling the GPU memory with the template data (both polarizations). \textbf{At this stage the generation is completed}. This function accepts some optional FLAGS used to control, for example, template normalization, template domain generation and host memory synchronization. The normalization code is based on  a distributed reduction algorithm\cite{CUDAsdk} that involve  GPU shared memory and CPU in order to speed up computation.
	\item[device\_complex\_mf] this function applies matched filtering formula in frequency domain between each polarization and the input data chuck $S$, using the \ref{eq:matchedfilter}. The two outputs of this step are called single polarization "correlator". 

	\item[device\_cufft\_C2R] this function applies inverse DFT on each previous  polarization correlator output, in order to go back in time domain. 

	\item[device\_madd2] this function combines together the two polarization correlators output using the \ref{eq:matchedfilter}. The output is a time domain vector $c[t]$, called "correlator output".

	\item[device\_maxfind] this function is used to look for "correlator output" values above threshold, than could be associated with true detection. This function is implemented using a  distributed reduction algorithm that involves GPU, GPU shared memory and CPU in order to speed up this process. The clusterization phase is made only by CPU. This chose permits us to hide this computing part behind the next iteration.  
\end{description}

\subsection{FFT and memory access}

The core of the matched filtering technique is the DFT algorithm, permitting to speed up the correlations computation. This algorithm has a numerical complexity of $O(Nlog(N))$ respect to the time correlation, which has complexity of $O(N^2)$.
NVIDIA provides a specific library to compute FFT on GPU, called CUFFT. In the context of Data Analysis, FFT is one of the most common and important algorithm, used for example for spectral analysis, correlation and convolution and other common tasks. For that reason, we developed a specific test to evaluate its performances, comparing CUFFT results with the same CPU implementation of the benchmark, using the famous fftw library \cite{FFTW}. At the moment, the main limitation of cufft is the support for single precision only, that could produce some numerical problems when the signal dynamics is getting higher. In order to reproduce the same conditions on both GPU and CPU architectures, the fftw library has been compiled in single precision, single thread with sse support\footnote{Usually, the fftw single precision gains the performance  of 30\% respect to double precision}. 

\begin{figure}[ht]
\centering
\includegraphics[width=4in]{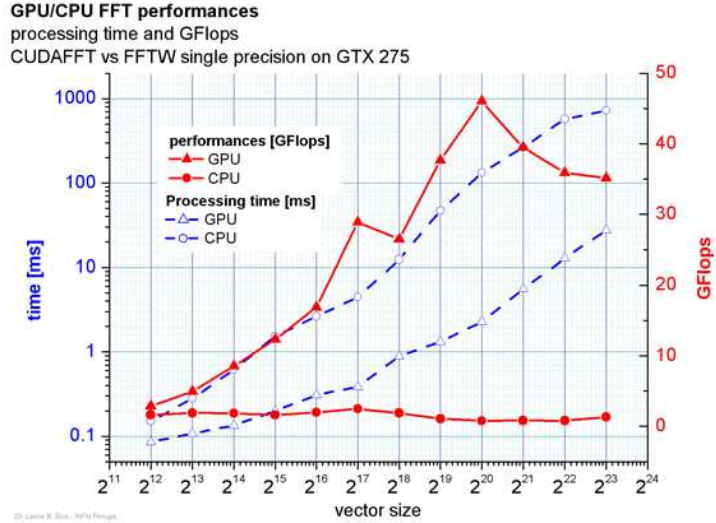}
\caption{We report the performances comparison between FFTW and CUFFT library. We build a specific test that measure the execution time and GFlops of the two implementation while vector length change. In the CB analysis usually we work arount vector of length $2^{19}$, where CUFFT gets a x50 performance gain.}
\label{fig:FFTWCUDAcomparison}
\end{figure}

The benchmark has been designed to record FFT execution time on  different size vectors, as show in figure \ref{fig:FFTWCUDAcomparison}. In order to get statistic, the code loops the FFT function over the same vector length\footnote{The number of loops depends on the vector size. For example, for a vector of length $4194304$, the number of loops per vector is $2000$ and that number change inversely proportional with the vector length.}.
In figure \ref{fig:FFTWCUDAcomparison} we report the comparison between cudaFFT(GPU-triangle) and fftw(CPU-circle) performances, measuring the average processing time (dash line) and the GFlops (line). As defined in fftw documentation \cite{FFTW} in real\-to\-complex case, GFlops are estimated as:
\[5 N log_2(N)\/\Delta t \/ 2\]
where "$\Delta t$" is the average FFT computing time expressed in microseconds.
For the cufft library case we use $5 N log_2(N)/\Delta t$ formula, because at the moment cudafft doesn't support an optimized routine for real to complex transformation. In fact the library function hides a complex to complex transformation. 

Obtained results show how, for a given vector length, the fft on GPU is many tens time faster than the fftw implementation. In CB signal detection analysis,  vectors with a length greater than $2^{19}$ samples are usually considered interesting. In this case the  performance gain is about $\mathbf{55}$ and the peak performances is obtained for vector length of $2^{20}$ where the GPU provides more than 45 GFlops of computing power.

\begin{figure}[ht]
\centering
\includegraphics[width=3.5in]{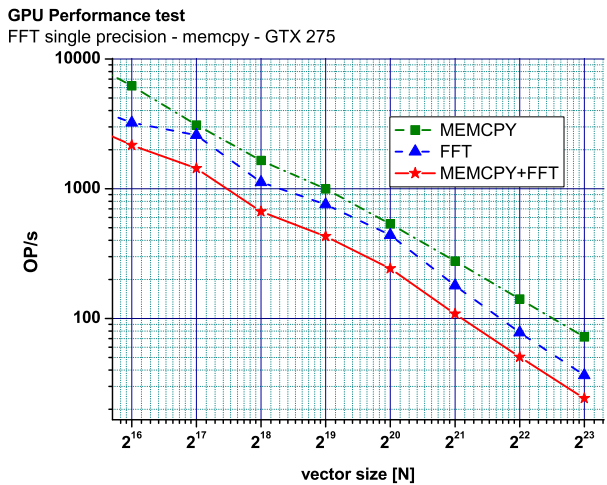}
\caption{In this plot we report the number of operations per secconds. The considered operations are: host-to-device memory copy, FFT alone and the combination of the two operations, simulating the overhead due to IO. These results show how the GPU has to be treated as a real stand-alone system and not as a co-processor, working as match as possible on GPU side and avoiding un-hidden IO}
\label{fig:FFTMEM}
\end{figure}

A remark, these high performances are achievable only if vectors data reside already on GPU device memory, otherwise data need to be first moved from host memory to device memory, paying a not negligible cost in terms of IO and computation delay. By that we evaluated with a benchmark the \textit{Host memory} - \textit{Device memory} communication overhead, that technically depends mainly on PCI-E bus speed. For example, in our signal detection case, we can encounter IO operation during the startup/init activity, reduction function and correlators maximum handling. We explored this aspect, estimating the IO transfer time changing the vector data length exchanged between the two memories. These results are reported in figure \ref{fig:FFTMEM}
where we show the number of copies per second(square symbol) respect to different vector data lengths. Moreover we show the equivalent number of FFT per second (triangle symbol)  applied on the same data, without memory copy (these are the same results used for \ref{fig:FFTWCUDAcomparison}) and the combination of Memory Copy and FFT algorithm (star symbol). These results provide the FFT performances that can be achieved when FFT is applied on data loaded  each time from host memory to Device memory. As effect the memory copy step produces a performance loss of $\mathbf{40\%}$ respect to a pure FFT computation. By these results we can make a consideration about GPU programming style,  learning how using GPU just as a co-processor, for example replacing fftw library functions with CUFFT library function , it is not in principle the best design. In order to get top performances from GPU architecture it is important to organize the code having an high computing GPU density and writing GPU oriented code. The work reported on this paper follows these rules, applied to a fully GPU based detection pipeline.   

\subsection{template generation and normalization}

The first step of the detection procedure is the \textbf{template generation} and \textbf{normalization}. 

\subsubsection{template generation}
The our knowledge about coalescing binaries signal is only theoretical and the waveforms are  only available with an approximation regime. The Post Newtonian \cite(GWapprox2,GWapprox3) approximation of order $2.5$ in phase and $0$ in amplitude has been used in the current version of our prototype library, but  there are no particular limitations on implementing other generators with higher accuracy level. The chosen approximation is enough for the purpose of this paper, because it permits to have a speed up factor between GPU and CPU that, at first order,  can be considered roughly constant respect to the approximation order. 

Looking to figure (\ref{fig:pipeline}), the processing steps implementing template generation and normalization are: \textbf{cuInspiral\_host\_init}, \textbf{cuInspiral\_device\_init}, \textbf{cuInspiral\_device\_template\_proc()}. 

The most computing demanding part is the template processing, where the two polarization of the template signal are generated, FFT transformed and normalized.

\begin{figure}[ht]
\centering
\includegraphics[width=3.5in]{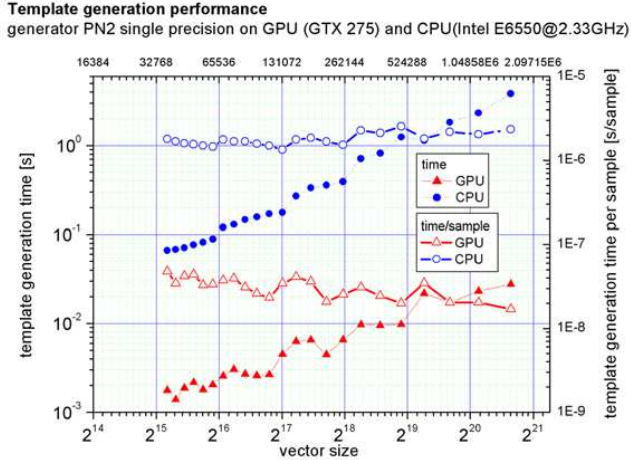}
\caption{In this plot we report the performances comparison between GPU and CPU about coalescing binary signal generation. With this benchmark we measure a gain of two order of magnitude in the region close to vector size=$2^{20}$. The generation used is a PN 2.5 in phase and zero in amplitude. In general we think to obtain the same gain factor also usign other more accurated generators.}
\label{fig:generatore}
\end{figure}

\textbf{The normalization} is the procedure used to define a template scale factors in such a way that  $<\theta_+,\theta_+>=1/2$ and $<\theta_\times,\theta_\times>=1/2$. This operation has been implemented in GPU with a complex product and reduction algorithms, which uses GPU shared memory to speed up the computation and the CPU as final accumulator. In detail we uses device shared memory to speed up the  $N=\sum_N res[i]$ operation. CUDA provide a fast shared memory region of 16KB for each multiprocessor\cite{CUDA} that can be shared among threads running on the same multiprocessor. This can be used as a user-managed cache and moreover it is much faster than global memory of the GPU device, permitting higher bandwidth operations. To apply such solution to our problem we implement a reduction procedure where we divide the computation in sub-domain  $N_s=\sum_{N/S} data[i]$, which result is stored into shared memory as first step, and than copied into host memory, performing the final $N=\sum_S N_s$. For more information about this technique please refer to CUDA SDK\cite{CUDAsdk}. For the FFT we use CUFFT, as previously described and profiled.
Here we report  performance details about the kernels in charge of generating the gw waveforms. In figure \ref{fig:generatore} we show the benchmark results, where the signal length is reported in the horizontal axis and  the generation time (filled markers) and generation sample(empty markers) per second are reported in the vertical axis. The test measures the average time spent on signal generation for the  two polarizations $h_+^i$,$h_\times^i$ for both architectures, GPU and CPU\footnote{A CPU version of the code has been written.}.
Considering the range reported in figure (\ref{fig:generatore}), GPU implementation provides a gain factor of \textbf{x[80-120]}, where  performances increase  for longer vectors. This behavior can be better observed looking  on the normalized curves, representing the number of signal samples generated  per second. In the CPU  case performances are roughly constant while vector size increases, but in the GPU case  the execution time per sample decreases.

GPU implementation of GW signal generators permits to change perspective about  gw signal generation and handling problem. In fact, actually the CPU based detection algorithms try to optimize this part essentially using bufferization strategies, avoiding the templates regeneration at each processing loop but introducing a lot of complexity in the code. 

Results, reported in this work about the signal generation, show a performance boost of two orders of magnitude, that can permit to change approach, trying a brute force procedure, namely \textbf{generating  on-the-fly the complete waveforms at each processing loop}. In this way we have margin to simplify the pipeline code significantly, but keeping anyway the big performance gain. 

\subsubsection{Numerical accuracy}
Another test has been performed about the numerical accuracy of the generation phase, comparing the single-precision of GPU generator respect to the CPU double precision counterpart. In detail, since the equation \ref{eq:matchedfilter} is used in signal analysis to compare two signals, we can use the same equation to estimate the mismatch between them. That can be also seen in another way, knowing that the \ref{eq:matchedfilter} is the filter that maximize the signal-to-noise ratio, it gives us information on how match energy we can extract and detect from the given signal. 
So, we use the matched filter definition (eq.\ref{eq:matchedfilter}), $<\theta_{Cpu},\theta_{Gpu}>$, to write the mismatch function:
\[\label{eq:mismatch}\Delta = 1-\frac{<\theta_{Cpu},\theta_{Gpu}> } {<\theta_{Cpu},\theta_{Cpu}>}\]
In case of two of perfectly identical signals normalized to $1$ , it gives  $0$ as result (no lose); while it gives a value $>0$ in all the other cases, where the output value is  proportional to the amount signal energy lost. 

We evaluate the signal generation numerical error, applying the mismatch function between GPU and CPU gw signal for the same CB source, exploring a parameter space in the range $[1-100]Ms$. \textbf{The measured mismatch error  between the two signals is lower than $0.03\%$}. In CB analysis usually a much higher incertitude of the order of few percent ($[1-5]\%$) is introduced by stochastic process and template space quantization. Thus, we can affirm that this is a largely tolerable error. 

Moreover, in order to evaluate the error due to FFT algorithm we perform the same comparison in frequency domain. In this case, obtained results report an higher error level of about $0.1\%$, imputable to FFT numerical error, but also in any case it is completely tolerable.

\section{GPU based detection pipeline performances and profiling}

In this section we report the pipeline performances and profiling analysis, in order to evaluate the performance of our prototype library and to get details about each used algorithm.  In some sense this section report the most important result of this work, because the processing time needed to complete the pipeline for each template processed is the main number of merit that can be use to make comparison (with eventually the appropriate correction and normalization) with other pipelines built with the same procedure.

The test has been produced, building up a benchmark over the GPU CB pipeline that generates a template grid uniformly distributed and applies such grid of templates to a given input, using the pipeline discussed in this paper. Timing information about each computing step are stored. 

The configuration of the simulation was:

\begin{itemize}
\item Intel E6550 2.33GHz
\item 2GB RAM DDR2 @800MHz
\item GPU board NVIDIA GTX275
\item PCI-E v2
\item Ubuntu 9 OS
\end{itemize}  

We report also some technical details about graphical board and the environment:

\begin{itemize}
\item 240 processing cores, distributed on 30 MultiProcessor with 8 cores each one.
\item Core\/Shader\/Memory clock rate of 0.6/1.4/2 Ghz
\item Memory Bandwidth of 127 GB/s (1GB GDDR3)
\item 1010 Declared GFlops (MADD + MULL)
\item CUDA driver version 2.3
\item CUDA capability version 1.3
\end{itemize}

Given a vector length of $2^{20}$, CB signal source $[1.4,1.4]Ms$ and sample rate of $4kHz$, the cuInspiral pipeline is completely processed in:
 pipeline time/template = \textbf{28ms} (average time)

This is a very impressive result, if compared to the actual pipeline performances.  Our prototype GPU CB pipeline gains more than a \textbf{factor 50}, providing a good idea about the implication of these new computing architecture on the GW analysis and physics.    

\begin{figure}[ht]
\centering
\includegraphics[width=3.5in]{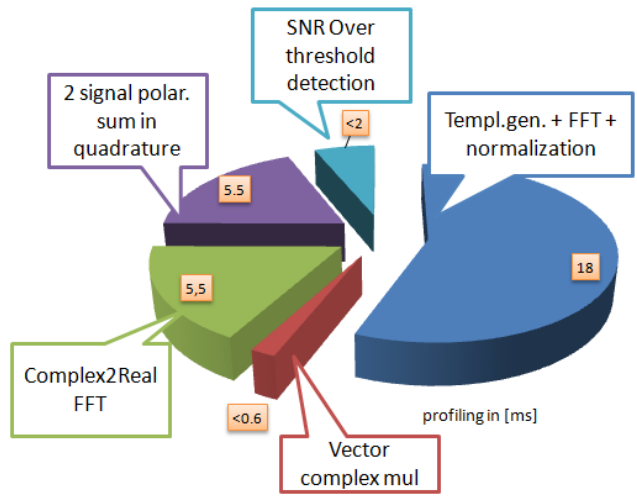}
\caption{In figure we report the pipeline profiling, reporting the execution time for each important section. These results have been obtained  with vector length of $2^{20}$.}
\label{fig:profiling}
\end{figure}

In more detail, we report the CB pipeline profiling, shown in figure \ref{fig:profiling}. By that we can make some considerations. The first is that the processing phase \textit{\textbf{template on-the-fly generation} + \textbf{FFT} + \textbf{normalization}} occupies roughly $55\%$ of the pipeline time. The second consideration is that the reverse FFT and the correlator computation (made adding in quadrature the two partial correlators), weigh similarly for the $15\%$ of the time.

Vector complex multiplication of the matched filtering formula and the correlator output above threshold detection are less critical, with an impact in the whole time of roughly $5\%$. 

\begin{figure}[ht]
\centering
\includegraphics[width=3.5in]{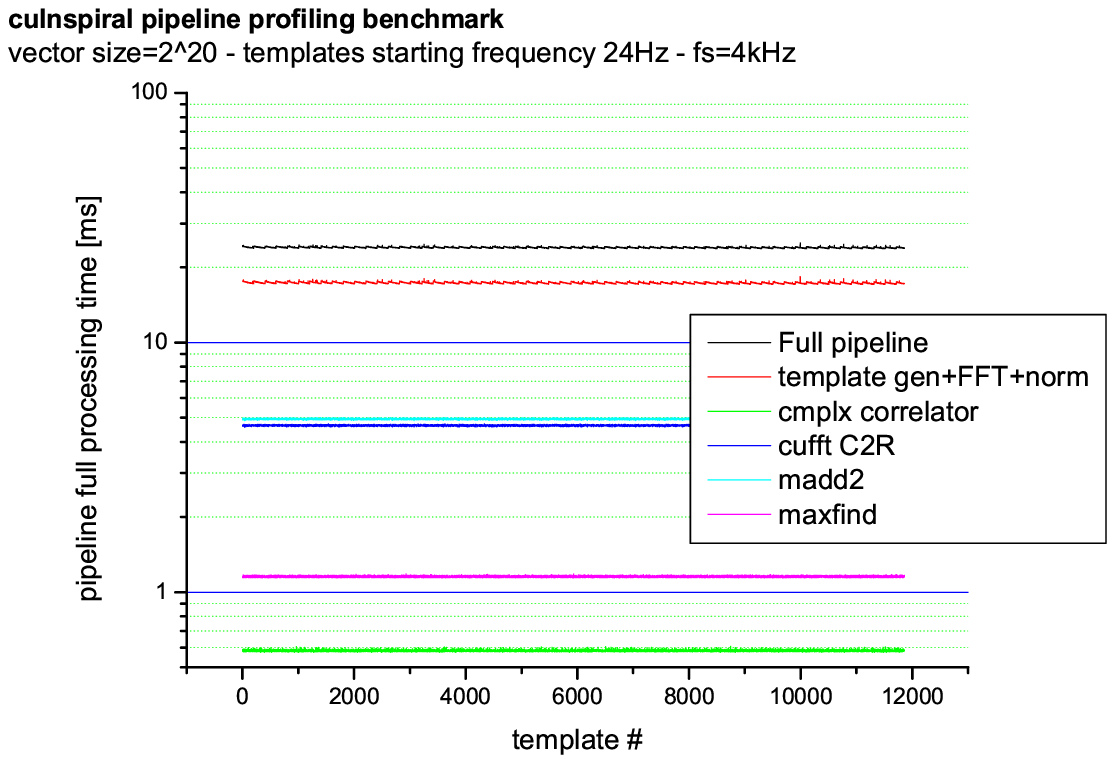}
\caption{In this plot we report the benchmark made to evaluate the stability of the code/hardware. In this case we run the pipeline over a very long run and we record the ordinal number of the template loop. At this level we not record any strange behavior. We have just to report a temperature increase of 35 celsius degree (up to 85 degree)}
\label{fig:pipelineprofiling}
\end{figure}

We made a test also to explore  the stability of the computation for long run.
This had been done keeping the vector length fixed to $2^{20}$, and processing a template bank with more than $1000000$ elements and recording the timimng informations. Results are shown in figure \ref{fig:pipelineprofiling}, where we report only the first 12000 loops.  
\begin{figure}[ht]
\centering
\includegraphics[width=3.5in]{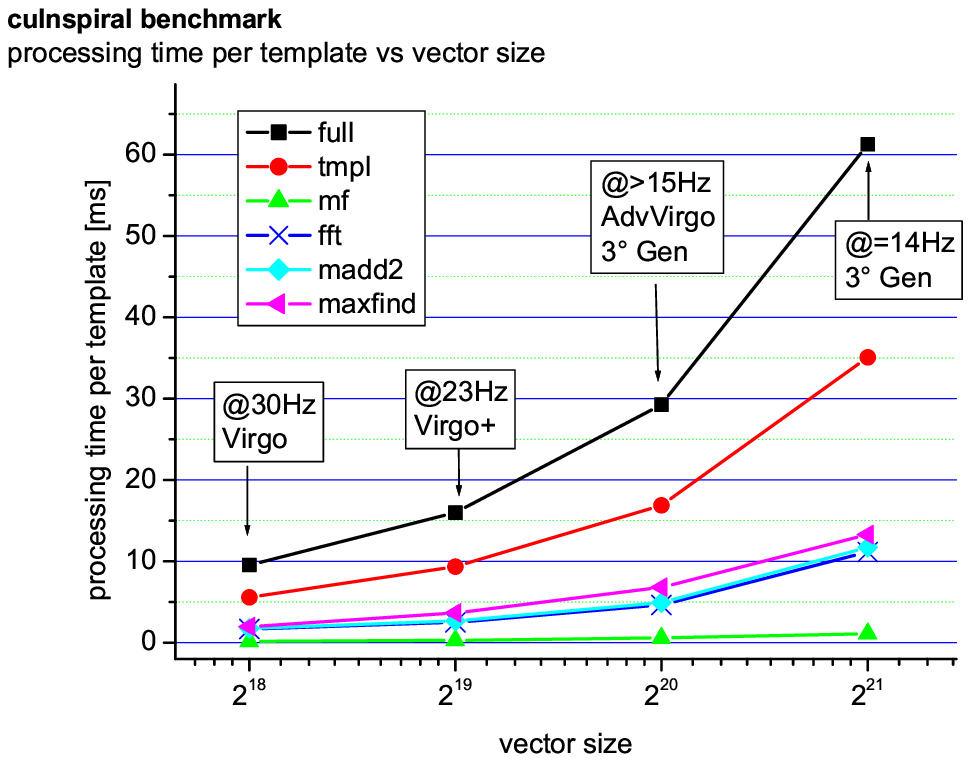}
\caption{In this plot the whole and profiled pipeline execution time is reported respect to the vector data size. A different vector length can be roughly associated to a "region of interest" for different gravitational wave detector. We highlight Virgo, LIGO (I gen), AdvVirgo, AdvLIGO (II Gen), Einstein Telescope (III Gen)}
\label{fig:pipelinepervectorsize}
\end{figure}
Results report a very stable behavior with time. During the test GPU temperature increases from $51$ to $84$ celsius degree  during  the first 5 minutes of run. After it was quite stable. 

\subsection{Online analysis considerations}
In this subsection we use our results to report some evaluation and implication about a GPU implementation of a CB pipeline in the online analysis case. On doing that we estimated the maximum number of templates that a single GPU board can process requiring that the computation has to be fast enough to follow the online data acquisition. This investigation is made changing the template length of binary system of $[1.4;1.4]Ms$ modifying the low frequency cut-off\footnote{Lowest frequency cut-off correspond to longest template}.  Results are reported in table \ref{tab:cutof} where we chosen as low frequency cut-off: $[30,23,15,14]Hz$. Each frequency can be associated to different region of interest for different gravitational wave detector generations.
 
\begin{tabular}{ | l | l | l | l |}
\hline
Vector Size&Cut in frequency& max template N\\ \hline
$2^{18}$&30 Hz& 3435 \\ \hline
$2^{19}$&23 Hz& 4095 \\ \hline
$2^{20}$&15 Hz& 4478 \\ \hline
$2^{21}$&14 Hz& 4279 \\ \hline
\label{tab:cutof}
\end{tabular}

In figure \ref{fig:pipelineprofiling} we report similar information with profiling data and labeling the regions of interest. In particular we can observe how at $30Hz$ we can compute online $4000$ templates, that is the roughly number of templates involved in the  analysis of first detector  generation. This means that with one single board we can cover the actual detector online analysis requirements. This is possible thanks to the factor $50$ obtained with the GPU CB pipeline code.   
  
Lower frequency cut-of are interesting for advanced detector such as AdvVirgo and AdvLigo and for third generation such as Einstein Telescope. About the third generation, we have to recall that Einstein Telescope by design could use a low frequency cut\-off $1-2Hz$. Under these condition templates could be 5-6 days long, requiring obviously to change the standard matched filtering procedure. For example we can figure out to use a multi band - hierarchical approach, where we divide the frequency range in more bands, making first the matched filtering on the higher frequency band. In this region the signal is shorter and has the highest amount of SNR. After that "detection" phase, the other bands can be recombined to acquire and increase signal for the the parameters reconstruction phase. 
%
%
%
%
%
%
%
%
\section{Conclusions}
In this paper we report the first prototype pipeline for the coalescing binaries signal detection completely based on GPU code. Moreover the pipeline routines have been organized on library called cuInspiral that will be soon free released.   

Our work show how using GPU hardware solution such as NVidia CUDA/GTX275 we can gain at least a factor \textbf{factor 50} in performance respect to the similar CPU implementation. Obviously there are still a lot of margin of code optimization and moreover new innovative hardwares are under release by the producers. For example we expect using the new CUDA library and Fermi solution a \textbf{factor greater than 150}.

Now we are working on production version of cuInspiral where the code has been more optimized and provided of a more accurate GW signal generator and soon of a optimized chi-squared\cite{Chi2} test routine. Moreover a multi-GPU version of the code is under construction, permitting to exploit the computing power of systems mounting more than one GPU board.

In the perspective of the manycore computing and of the gravitational wave detectors of third generation, INFN in 2009 approved the experiment "Manycore Computing for future Gravitational Observatory" (MaCGO)\cite{MaCGO}\footnote{the writer of this paper is the MaCGO national deputy} with the aim of develop a numerical library using OpenCL on which work cuInspiral has an important role.

Our results demonstrate how the full GPU/manycore architecture programming can be very suitable for data analysis problem of second and, in particular, third generation of gravitational wave detectors, such as Einstein Telescope, providing a impressive computational power able to access physical problem previously inaccessible. The implementation of these new solutions is obviously not at zero cost, because many-core programming architecture needs a change of paradigm, but the effort is largely repaid with the achievable performances gain as it has been demonstrated in this work. Moreover we have to highlight how the many-core architecture is where the common CPU design are evolving. Thus, in next 10 years the actual gravitational wave data analysis (and maybe many other fields of physics) had to face this change if want to get power from moder architecture and solve the ever new and ever more computing demanding problems of Gravitational wave Physics.

\section{Acknowledgments}
The research leading to these results has received funding from the European Community's Seventh Framework Programme (FP7/2007-2013) under grant agreement n 211743 in the context of ET (Einstein Telescope)  Design Study project.
I wish to acknowledge also Dr. Michele Punturo for his valuable advices.

\end{document}